\documentclass[twocolumn]{aastex62}
\shorttitle{Ultra-Hot Jupiter Host Stars}
\shortauthors{Lothringer et al.}

\usepackage{amsmath}
\usepackage{xspace}
\usepackage{float}

\begin{document}
\title{The Influence of Host Star Spectral Type on Ultra-Hot Jupiter Atmospheres}
\author[0000-0003-3667-8633]{Joshua D. Lothringer}
\affiliation{Lunar and Planetary Laboratory, University of Arizona, Tucson, AZ, USA}

\author[0000-0002-7129-3002]{Travis Barman}
\affiliation{Lunar and Planetary Laboratory, University of Arizona, Tucson, AZ, USA}
\vspace{0.5\baselineskip}
\date{\today}
\email{jlothrin@lpl.arizona.edu}

\begin{abstract}

Ultra-hot Jupiters are the most highly irradiated gas giant planets, with
equilibrium temperatures from 2000 to over 4000 K. Ultra-hot Jupiters are amenable to characterization due to their high temperatures, inflated
radii, and short periods, but their atmospheres are
atypical for planets in that the photosphere possesses large concentrations of atoms and ions
relative to molecules. Here we evaluate how the atmospheres of these planets
respond to irradiation by stars of different spectral type. We find that ultra-hot Jupiters exhibit temperature inversions that are sensitive to the spectral type of the host star.  The slope and temperature
range across the inversion both increase as the host star effective temperature
increases due to enhanced absorption at short wavelengths and low pressures.
The steep temperature inversions in ultra-hot Jupiters around hot stars
result in increased thermal dissociation and
ionization compared to similar planets around cooler stars.
The resulting increase in H$^{-}$ opacity leads to a transit
spectrum that has muted absorption features. The emission spectrum,
however, exhibits a large contrast in brightness temperature, a signature that
will be detectable with both secondary eclipse observations and high-dispersion
spectroscopy. We also find
that the departures from local thermodynamic equilibrium in the stellar
atmosphere can affect the degree of heating caused by atomic metals in the
planet's upper atmosphere. Additionally, we further quantify the significance of heating by different opacity sources in ultra-hot Jupiter atmospheres.

\end{abstract}


\received{11 November 2018}

\accepted{27 March 2019}

\section{Introduction}
One of the defining characteristics of a hot Jupiter is that it is highly irradiated by a nearby host star. The radiation it receives from its host star largely determines the planet's atmospheric composition, structure, and circulation. While the first hot Jupiters were found around solar-type stars \citep[e.g.,][]{mayor:1995,butler:1997}, it has become clear that hot Jupiters exist around a wide range of host star types. For example, NGTS-1b orbits a M0.5 star (T$_{eff}$ = 3916 K, \citealt{bayliss:2018}), while KELT-9b orbits a A0-B9 star (T$_{eff}$ = 10170 K, \citealt{gaudi:2017}). 
Host star spectra vary greatly within this range of stellar types; not only do the stellar spectra peak at different wavelengths, but the very different atomic and molecular compositions of each star imprint different spectral features on a star's spectrum. For a given planet's equilibrium temperature, the irradiation a planet receives can vary greatly depending on what type of star it orbits.

The population of hot Jupiters that experience the most extreme irradiation, referred to as ultra-hot Jupiters, have equilibrium temperatures in excess of 2000 K and can experience irradiation several thousand times the flux received by the Earth from the Sun. Recent modeling has explored the unique physics and chemistry occurring in ultra-hot Jupiters, finding that thermal dissociation of molecules can result in an atmosphere dominated by atoms at the photosphere \citep{lothringer:2018b,kitzmann:2018,parmentier:2018,arcangeli:2018}. At such high temperatures, H$^{-}$ becomes a dominant continuous opacity source. Furthermore, absorption at UV and optical wavelengths by species like Fe can induce significant temperature inversions, even in the absence of TiO and VO \citep{lothringer:2018b}. The effects of different host star irradiation should be most important in this population of ultra-hot Jupiters, yet previous studies of host star irradiation were not optimized for the ultra-hot Jupiter regime \citep{molliere:2015}. 

In this paper, we investigate the effect that different irradiation spectra
have on a hot Jupiter's atmosphere by self-consistently modeling several hot
Jupiter scenarios, described in Section \ref{methods}. We explore each
scenario's resulting temperature structure (Section \ref{results:tps}) and
chemical composition (Section \ref{results:comp}). We then discuss the
implications of these results in Section \ref{section:discuss}.

\section{Methods}\label{methods}

We use PHOENIX \citep{hauschildt:1997,barman:2001} to calculate custom models for both the star and planet. PHOENIX calculates the atmospheric structure and composition in chemical equilibrium subject to the constraints of radiative equilibrium and hydrostatic balance. The models span the UV and IR from 10 to $10^6$ \AA, sampled every 0.1 \AA \xspace at UV, optical, and NIR wavelengths. The model includes opacity from 130 molecules and from atoms up to uranium. Sources for important molecular opacity data are shown in Table \ref{table:opac}. Models are constructed using 64 layers distributed evenly in log-space between $log_{10}(\tau) =$ -10 and 2, where $\tau$ is defined at 1.2 microns. This range roughly corresponds to pressures of 10$^{-10}$ bars and 0.25 bars for the stellar models and 10$^{-10}$ bars and 50 bars in the ultra-hot Jupiter models. We describe the stellar and planet models below.

 \begin{table}[t] 
	\centering  
	\caption{Sources for Important Molecular Opacity Data}
	\label{table:opac} 
	\begin{tabular}{p{1.9cm}p{5cm}}
		\hline
		Molecule & Source \\
		\hline 
		H$_2$O & \cite{barber:2006}  \\ 
				\hline 
		CO & \cite{goorvitch:1994}  \\ 
				\hline 
		CO$_2$ & \cite{rothman:2008}  \\ 
				\hline 
		TiO & \cite{schwenke:1998}  \\ 
				\hline 
		VO & Plez (Private Communication)  \\ 

	\end{tabular}
\end{table}

\subsection{Stellar Models} \label{methods:stars}

\begin{figure}[t]
	\center    
	\includegraphics[width=3.5in]{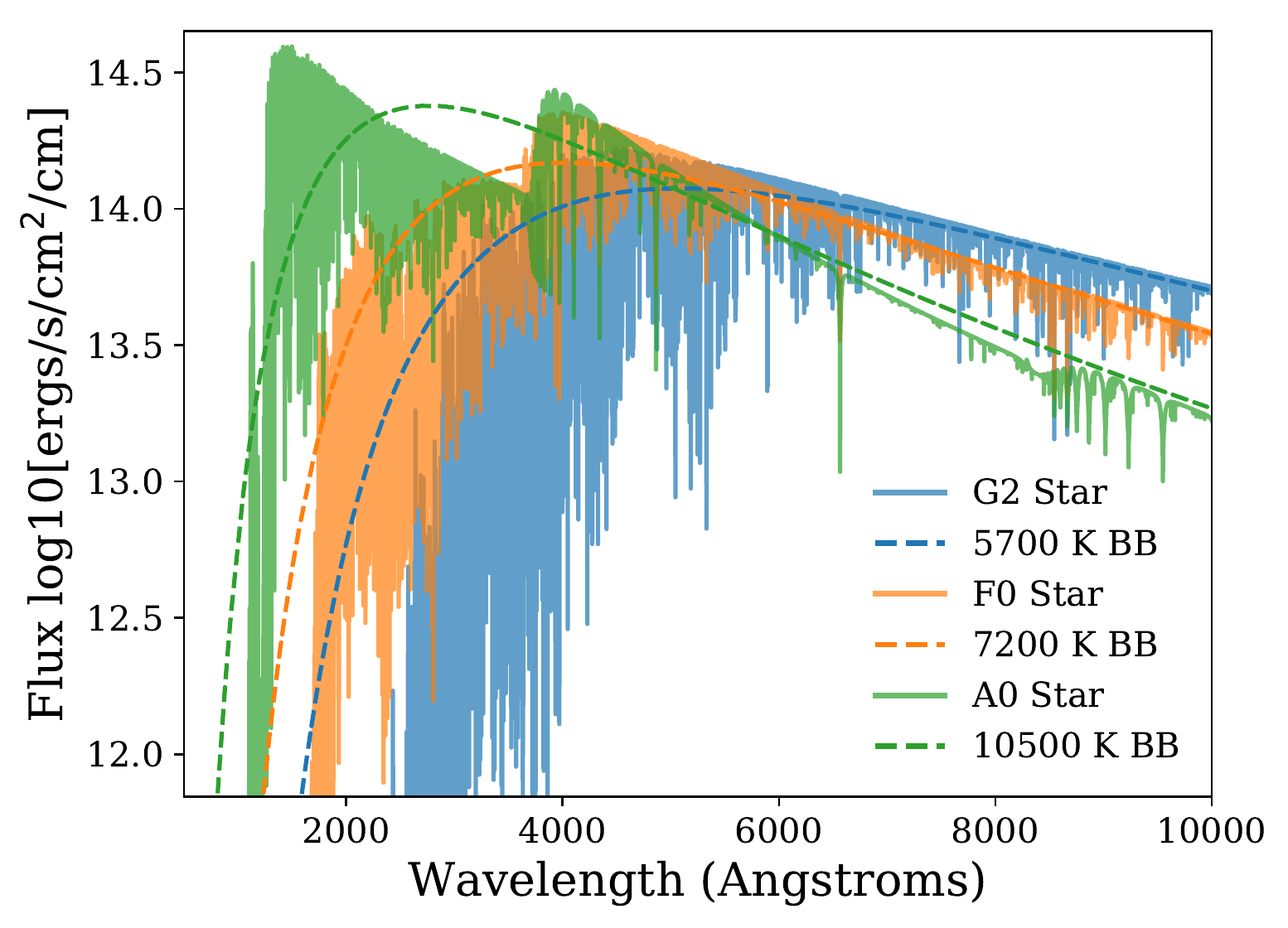}
	\caption{The host star fluxes at the location of a T$_{eq}$ = 3000 K planet used to irradiate our planet atmosphere models (solid) compared to blackbodies of the same T$_{eff}$ (dashed). While the total incoming stellar energy is equal, the A0 star (T$_{eff}$ = 10500 K) emits about half of its energy in the UV and has a prominent Balmer jump, while the G2 star (T$_{eff}$ = 5700 K) generally emits the vast majority of its energy at visible and IR wavelengths and exhibits deeper absorption lines. 
	\label{fig:stars}}
\end{figure}

\begin{figure}[t]
    \centering
    \includegraphics[width=3.5in]{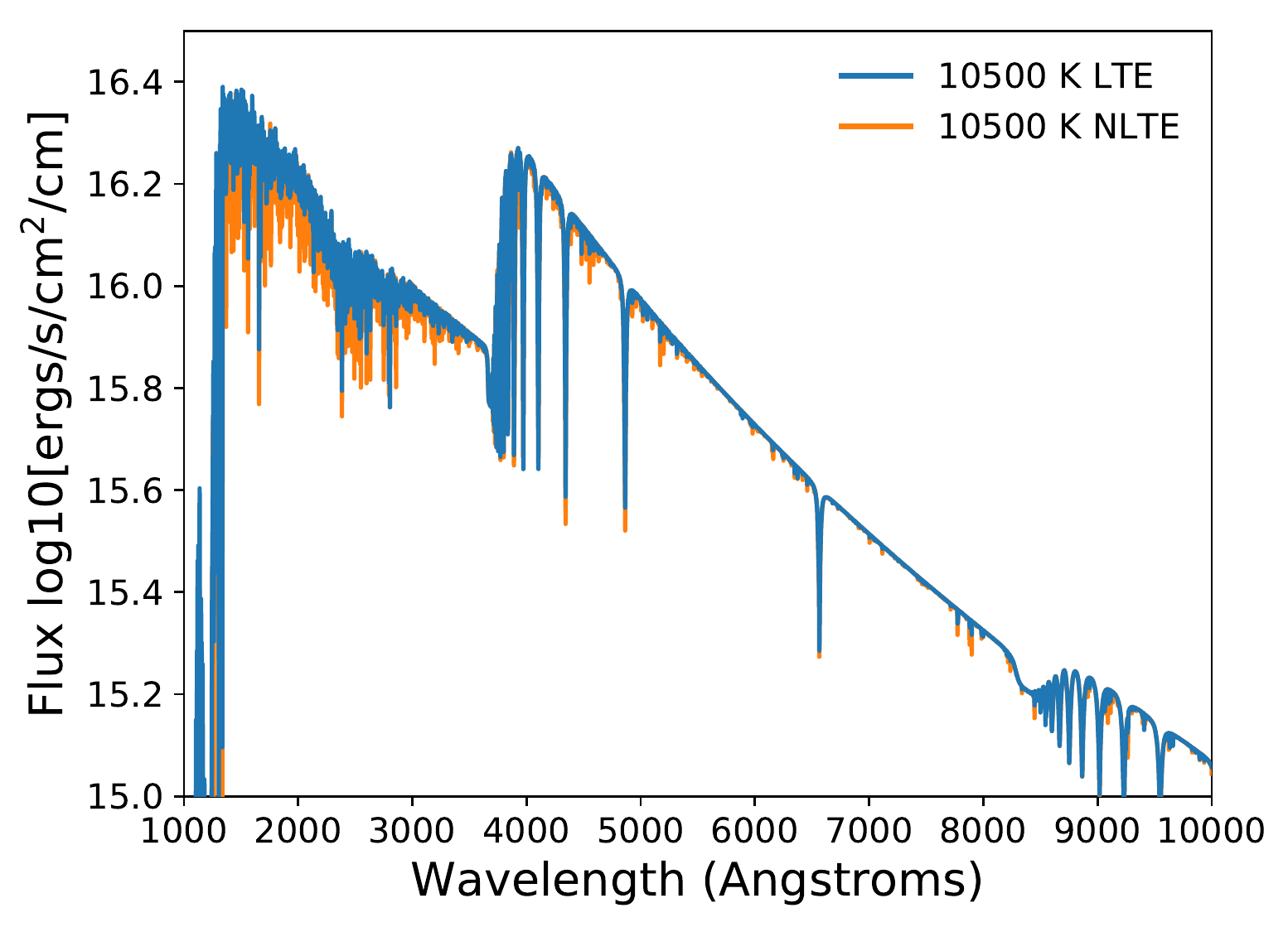}
    \caption{The host star flux of the A0 star in both LTE (blue) and when accounting for several atoms in NLTE (orange, see text) using the LTE temperature structure. Note that many spectral lines are considerably deeper in the NLTE case, leading to less flux absorbed in the upper atmosphere of the planet by species that absorb at those wavelengths.}
    \label{fig:nltestar}
\end{figure}

We use 3 different stellar types to explore the effect of host star irradiation on a hot Jupiter's atmosphere: G2 (T$_{eff}$ = 5700 K, R$_{star}$ = 1 R$_\odot$), F0 (T$_{eff}$ = 7200 K, R$_{star}$ = 1.55 R$_\odot$), and A0 (T$_{eff}$ = 10500 K, R$_{star}$ = 2.34 R$_\odot$). In both effective temperature and radius, these stars correspond to Sun-like, WASP-33-like \citep{stassun:2017}, and KELT-9-like \citep{gaudi:2017} host stars, respectively. We choose not to include any host stars much cooler than G2 because it is unclear whether a planet could survive close enough to such a star for the equilibrium temperatures we associate with ultra-hot Jupiters. Each model has solar metallicity. 

Figure \ref{fig:stars} shows the different stellar spectra used to irradiate the planet models, scaled to the location of a T$_{eq}$ = 3000 K planet such that the wavelength-integrated flux is the same. 
Beyond the fact that the hotter stars emit more short-wavelength flux due to Planck's law ($\sim$50\% of an A0 star's energy is emitted in the UV, compared to $\sim$5\% for a G2 star), the stellar spectra also vary due to the Balmer jump and differences in line absorption. The Balmer jump is caused by bound-free absorption by atomic hydrogen's second energy level, resulting in a increase in flux at 3646~\AA \xspace with the largest jumps generally found in {A-type} stars \citep{bessell:2007}. Line absorption can also vary with stellar type depending on the species present in a star's atmosphere. Atomic metals are particularly important for the ultra-hot Jupiters we consider here, because absorption of irradiation by metals in the planet's atmosphere can drive significant heating \citep{lothringer:2018b}. However, the magnitude of this heating is determined by the amount of flux emitted by the star at the wavelengths those metals absorb. The depth and width of atomic metal absorption lines in host stars can therefore mediate the level of heating in hot Jupiter atmospheres.

As such, departures from local-thermodynamic equilibrium (LTE) will affect line depths and widths in hot stellar atmospheres. At temperatures of several thousand Kelvin, radiative rates begin to become important relative to collisional rates in determining the atomic and molecular level populations, driving the level populations far from the Boltzmann distribution, which governs the populations of levels in LTE. We therefore model the stars in non-local thermodynamic equilibrium (NLTE). 
PHOENIX accounts for NLTE effects by self-consistently solving the multi-level rate equations \citep[e.g.,][]{hasuchildt:1995}. The NLTE models consider H I, He I-II, C I-IV, N I-IV, O I-IV, Mg I-III, and Fe I-IV in NLTE. Figure \ref{fig:nltestar} shows the difference between the KELT-9-like A0 (T$_{eff}$ = 10500 K) stellar model in LTE versus the same model in NLTE for the same temperature structure and composition. In particular, the Fe lines in the star are significantly deeper in NLTE. We explore how this behavior affects an ultra-hot Jupiter atmosphere in Section \ref{section:nlte_tp}.

We do not model a chromosphere for simplicity and because the XUV radiation coming from the chromosphere does not significantly affect the atmosphere at the pressure levels we focus on here, namely pressures greater than 1 $\mu$bar. Stellar chromospheres are, however, important to consider for the planet's upper atmosphere and for atmospheric escape. The existence or non-existence of a stellar chromosphere can determine whether a planet is in an energy-limited escape regime or a Jeans escape regime (Fossati et al. 2018). In this work, we focus on the impact of differences in the FUV, NUV, and visible flux coming from the stars rather than the XUV radiation.

 \begin{table}[t] 
	\centering  
	\caption{Planet Model Properties}
	\label{table:properties} 
	\begin{tabular}{p{1.9cm}p{1.8cm}p{1.8cm}p{1.8cm}}
		\hline
		Equilibrium Temperature (K) \tablenotemark{1} & Host Star Effective Temperature (K) & Host Star Radius (R$_\odot$) & Orbital Distance (AU) \\
			\hline 
		2250 & 5700 & 1 & 0.021  \\ 
		\hline 
		2250 & 7200 &1.55& 0.052  \\ 
		\hline 
		2250 & 10500 &2.34& 0.168  \\ 
		\hline   
		3000 & 5700 &1 & 0.012  \\ 
		\hline    
		3000 & 7200 &1.55 & 0.029  \\ 
		\hline 
		3000 & 10500 &2.34 & 0.094 \\ 
	
	\end{tabular}
	\tablenotetext{1}{These equilibrium temperatures and properties correspond to dayside-only heat redistribution since ultra-hot Jupiters are predicted to have poor heat redistribution \citep{tad:2016,tad:2017}.}
\end{table} 

\subsection{Planet Models}

We use the same methods as \cite{lothringer:2018b} to model the planetary atmospheres. We assume the planet has dayside-only heat redistribution and construct the planet models for two equilibrium temperatures: 2250 K and 3000 K. These parameters and their corresponding orbital distance are listed in Table \ref{table:properties}. The incoming stellar light is assumed to be isotropic. Additionally, the models assume an internal heat flux of $\sigma T_{int}^4$, where T$_{int}=125$ K, which has a negligable effect on the resulting temperature structures because of the high level of irradiation. Also note that the high irradiation experienced by ultra-hot Jupiters pushes the radiative-convective boundary to $\tau > 100$, outside of the upper optical depth boundary of the models. We therefore turn off convection to speed up convergence, but also make sure the final models are non-convective.

In addition to the continuous opacity sources listed in \cite{lothringer:2018b}, we also add molecular bound-free opacity from H$_2$, CH, NH, OH, and CO \citep{kurucz:1987,chan:1993,visser:2009,fontenla:2015}. These opacity sources have a minor effect on the atmosphere, likely due to those molecules' low abundance at high temperatures and low pressures. These opacities may become more important in cooler atmospheres. Our fiducial model is also the same as in \cite{lothringer:2018b}: a 1 Jupiter-mass, 1.5 Jupiter radius planet with dayside temperature redistribution and solar metallicity assumed. 

As in \cite{lothringer:2018b}, we assume LTE in our planet models. NLTE effects are likely important in some parts of the atmosphere of such highly irradiated planets, as radiative rates of excitation may approach and exceed collisional rates of excitation. Our results presented here are likely to be most accurate for pressures below a $\mu$bar, where collisions can remain dominant, keeping the atmosphere closer to LTE. We will explore NLTE effects in ultra-hot Jupiter atmospheres in future work.

\section{Results} \label{results}

\subsection{Temperature Structures} \label{results:tps}

\begin{figure*}[ht!] \hypertarget{fig:tps}{}
	\gridline{\fig{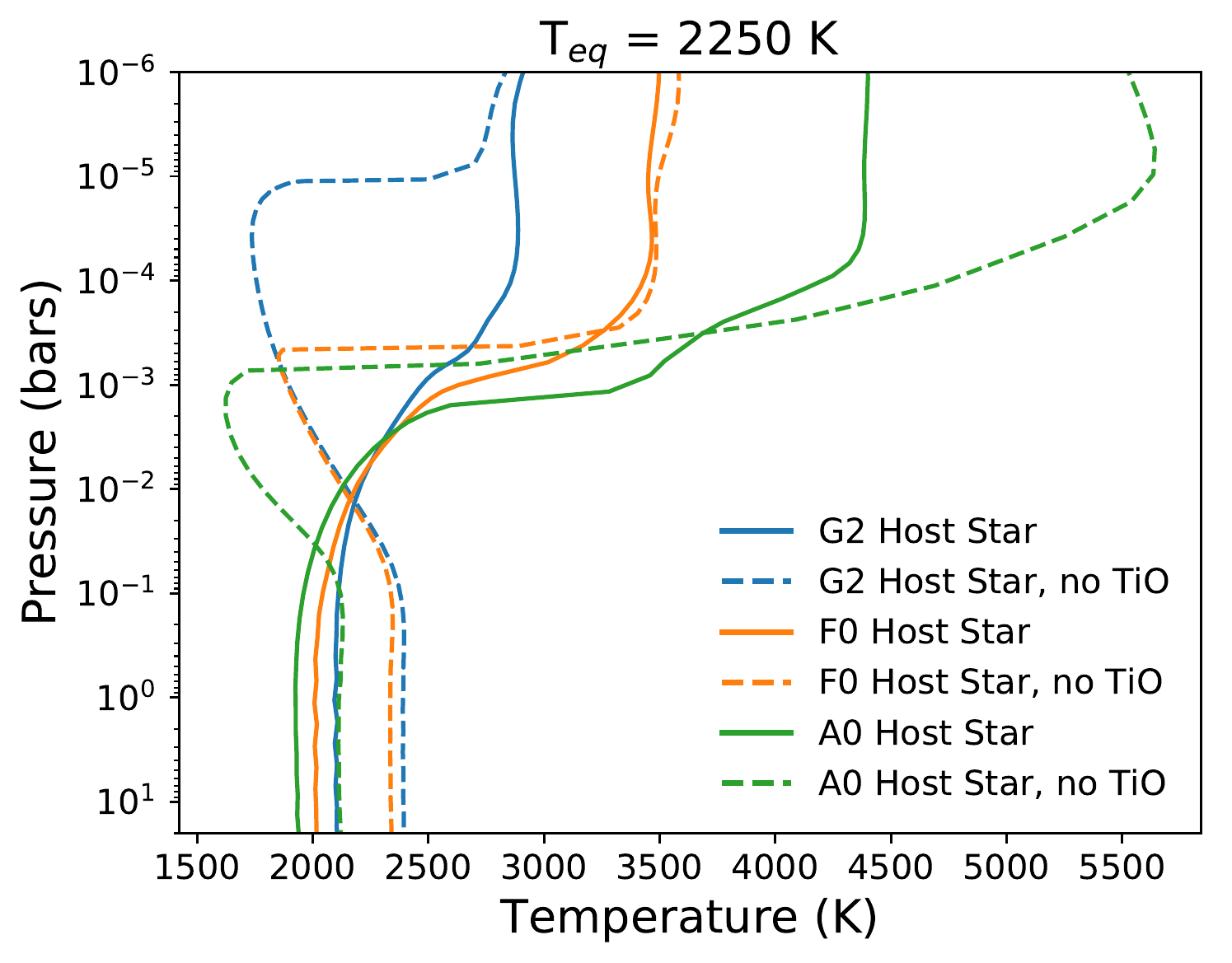}{0.5\textwidth}{(a)}
		\fig{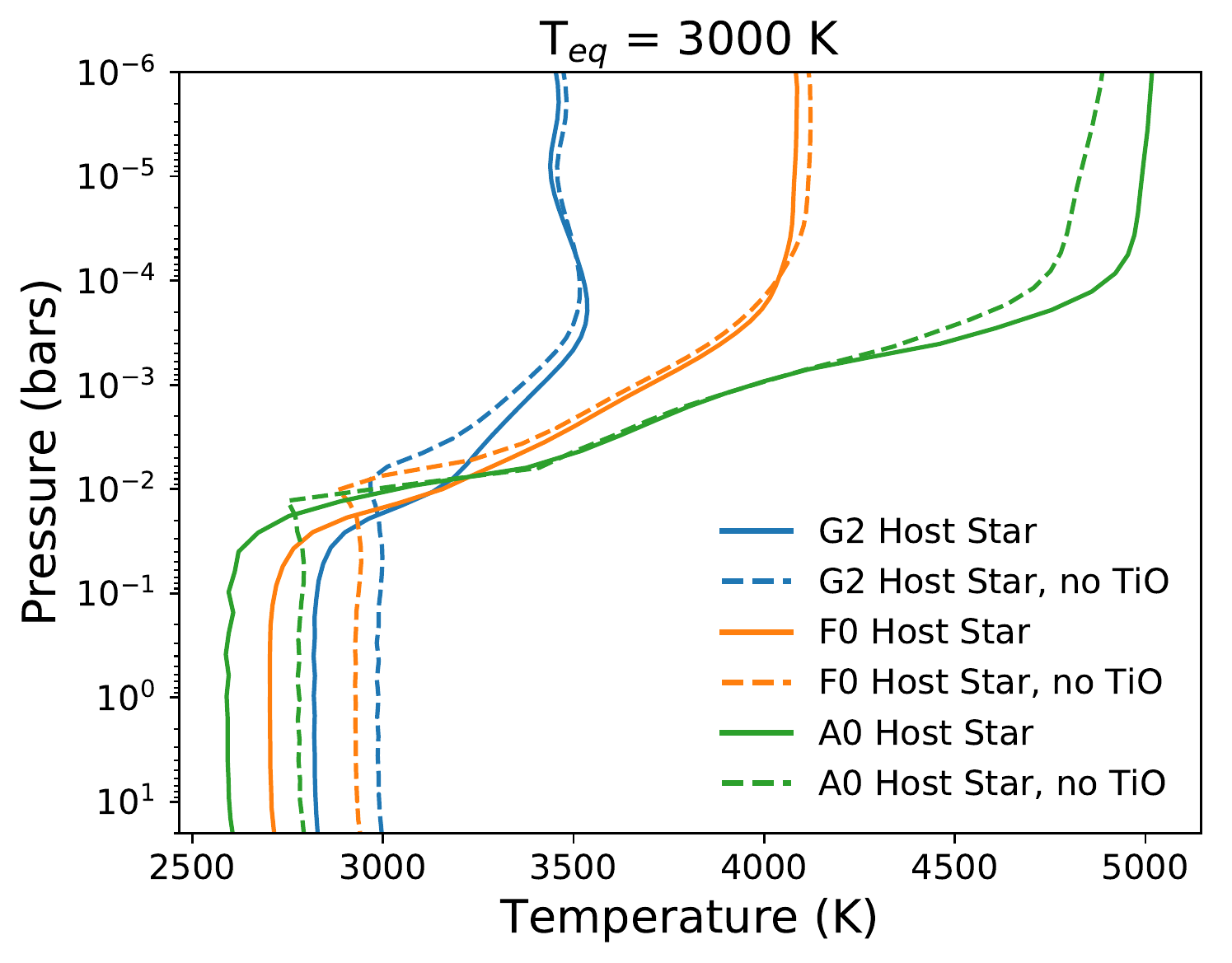}{0.5\textwidth}{(b)}}
	\caption{Pressure-temperature profiles of planets around G2, F0, and A0 host stars, corresponding to T$_{eff}$ = 5700 K, 7200 K, and 10500 K, respectively. The figure at left shows a Jovian planet with T$_{eq}$ = 2250 K and the right figure shows a Jovian planet with T$_{eq}$ = 3000 K, with and without TiO and VO.}
\end{figure*}

\begin{figure*}[ht!]\hypertarget{fig:opac}{}
    	\gridline{\fig{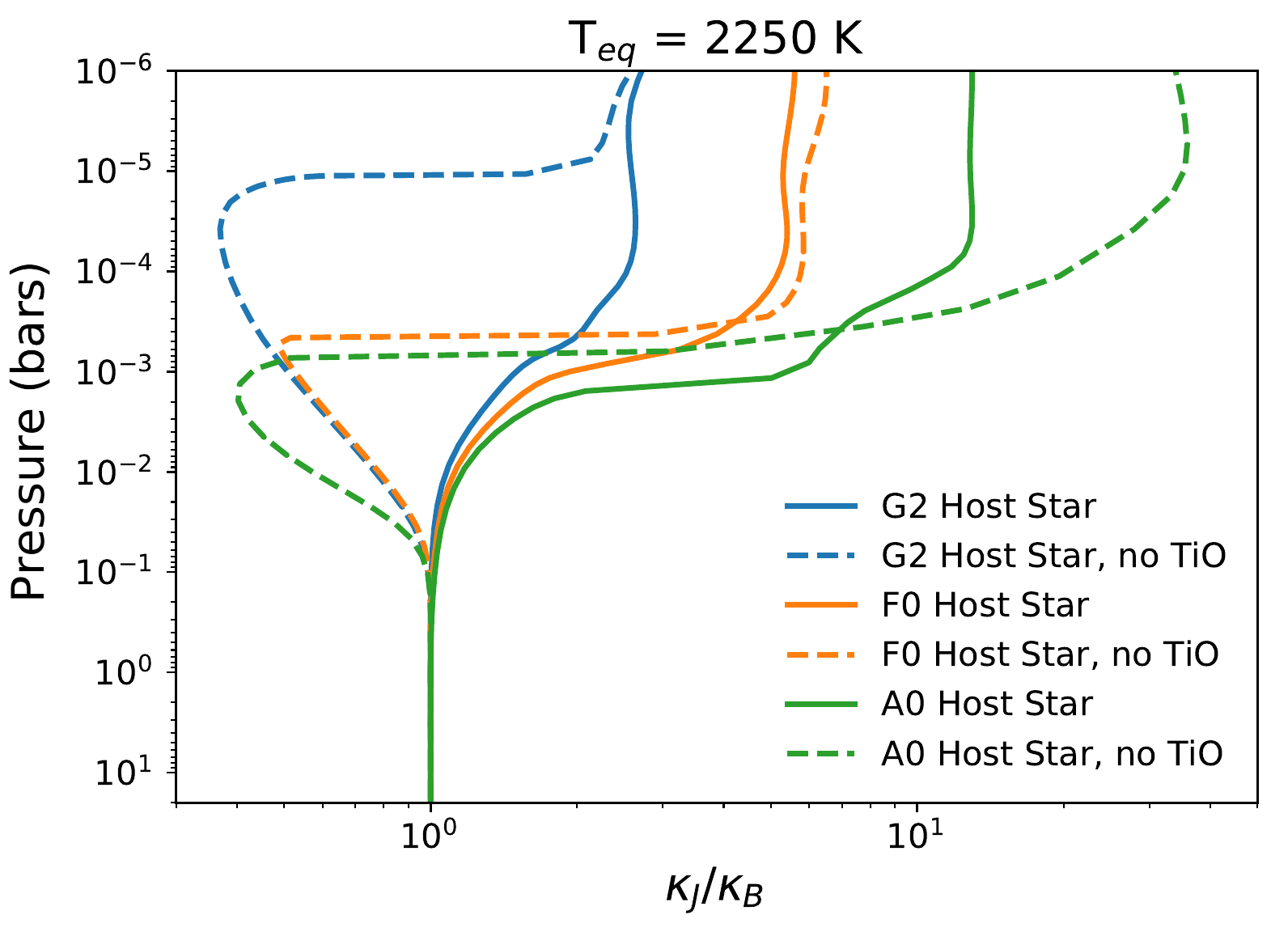}{0.5\textwidth}{(a)}
    		\fig{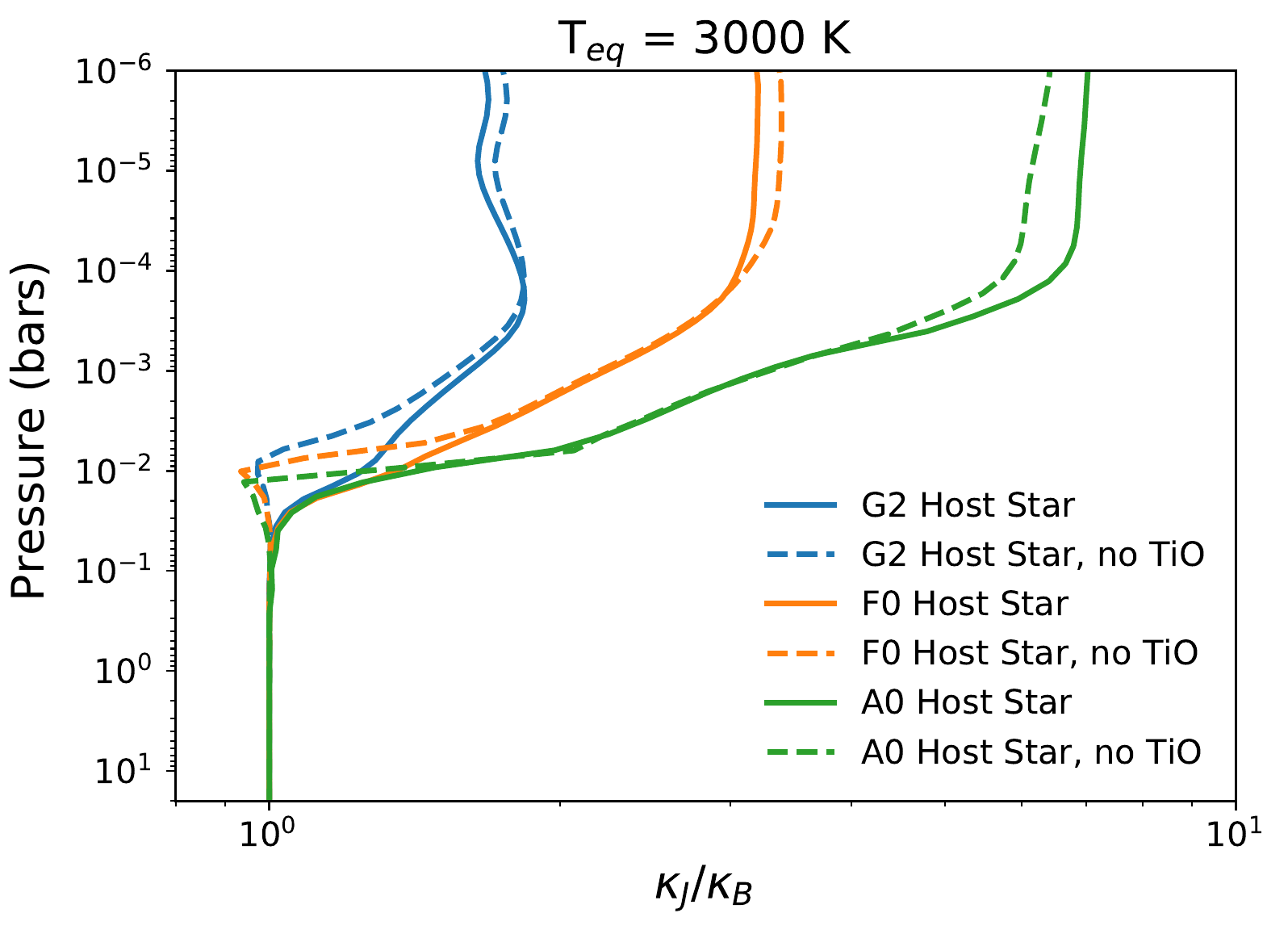}{0.5\textwidth}{(b)}}
    \caption{The ratio of the absorption mean opacity to the local Planck mean opacity for each of the models shown in Figure~\hyperlink{fig:tps}{3} (see text for details).}
\end{figure*}

The temperature structures of ultra-hot Jupiters, in general, have steep temperature inversions beginning at about 10-50 mbar, as shown in Figure~\hyperlink{fig:tps}{3}. The temperatures at 1 mbar are in most cases ${>}{500}$ K higher than the temperatures at 100 mbar. The properties of the inversion are
regulated primarily by the absorption of short-wavelength (${<}0.5$ \AA)
irradiation by atomic metals
\cite[and see below]{lothringer:2018b}. The inversion, however, is also sensitive to TiO and VO
for T$_{eq} \leq 2500$ K, where even modest mixing ratios of these molecules can
induce inversions at the observable photosphere
\cite[]{hubeny:2003,fortney:2008}.  Even without TiO and VO, temperature inversions
will form, albeit at pressures less than 1 mbar and depths not well probed by
low-resolution near infrared observations. Dayside spectra of T$_{eq} = 2500$ K hot Jupiters without TiO and VO will therefore exhibit absorption features rather than emission features.

In both the T$_{eq}$ = 2250 K and 3000 K cases, the slope and maximum temperature of
the inversion are strong functions of host star effective temperature.  For
example, in the T$_{eq}$ = 3000 K models, the planet with the KELT-9-like
T$_{eff}$ = 10500 K host star is over 500 K hotter at 1 mbar than the planet
with the Sun-like T$_{eff}$ = 5700 K host star. Interestingly, the base of
the temperature inversion occurs at similar pressures regardless of the host
star's type in most cases. For example, in all the T$_{eq}$ = 3000~K models, the temperature inversion begins at about 50 mbars. In contrast, the base of the inversion 
in the T$_{eq}$ = 2250 K models without TiO and VO do have a strong dependence on host star type.

Our models indicate that there exists some scenarios where a planet without TiO and VO will have an observable temperature inversion when it is around an early-type host star, but will not have a temperature inversion if it is around a cooler host star. In the latter situation, the irradiation does not have enough short-wavelength flux to drive sufficient heating in the atmosphere to fully invert the temperature structure at the photosphere. This bifurcation in atmospheres without TiO and VO occurs between the T$_{eq}$ = 2250 K and 3000 K. Predictions for whether a planet has an observable inversion or not in this equilibrium temperature range should take into account the host star's spectral type and the possible presence of TiO and VO.

The general behavior of the temperature structure at lower pressures is reversed at higher
pressures, where the ultra-hot Jupiters around hotter host stars have cooler temperatures
at pressures greater than 10 mbar when compared to similar planets around cooler stars. This is because of an anti-greenhouse effect. Ultra-hot Jupiters with the
steepest inversions will have absorbed most of the incoming stellar irradiation
in the inversion layers, resulting in less heating in the deep atmosphere and
therefore lower temperatures.  Note that even in the T$_{eq}$~=~2250 K model
without TiO (and therefore with a temperature inversion well above the near-IR
photosphere) the amount of short-wavelength irradiation can still influence the
temperatures at pressures above 1 mbar and across most of the observable
photosphere.

\subsubsection{The Effect of Opacity on the Temperature Structure}\label{opactemp}

\begin{figure*}[ht!]\hypertarget{fig:kj}{}
	\gridline{\fig{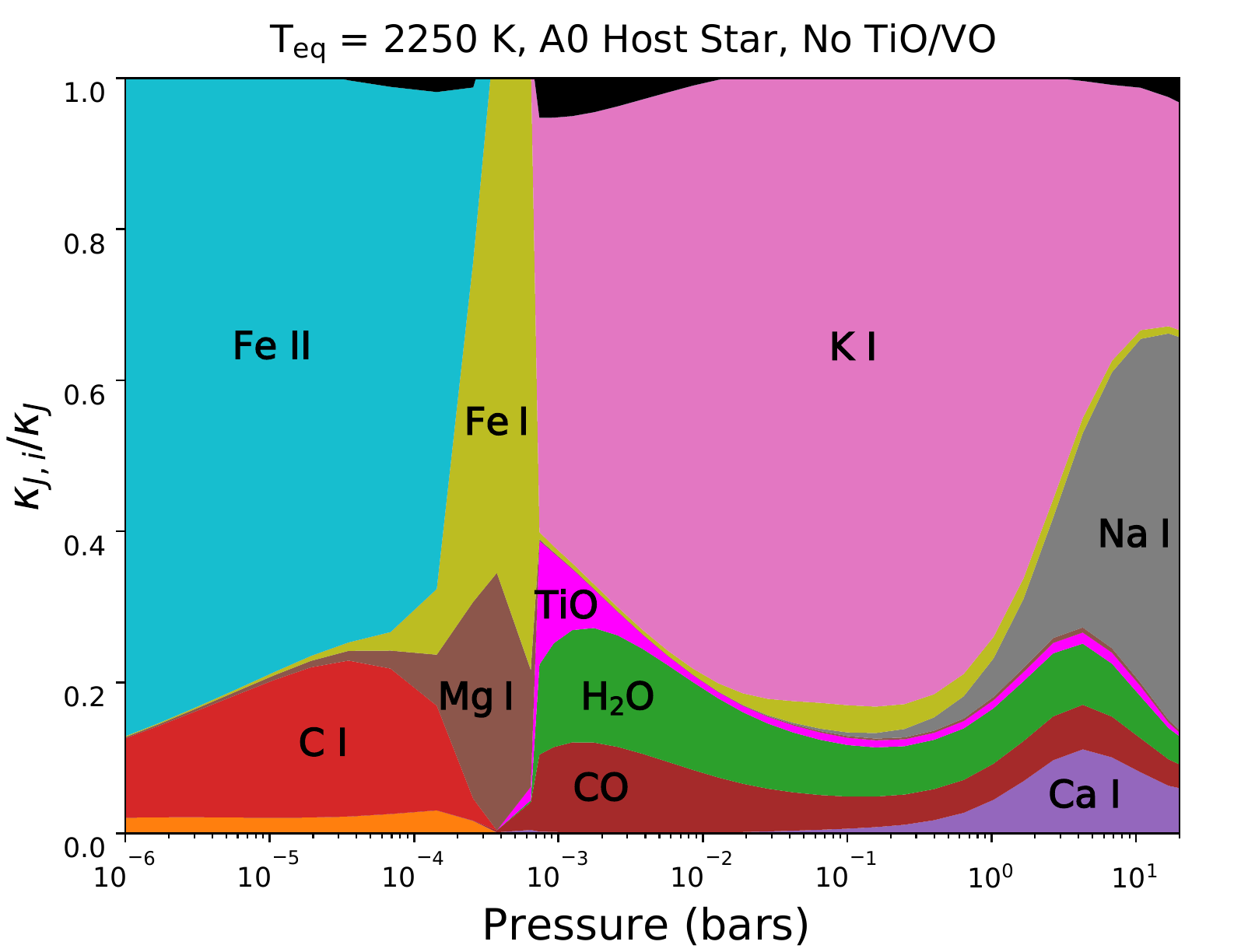}{0.43\textwidth}{(a)}
		\fig{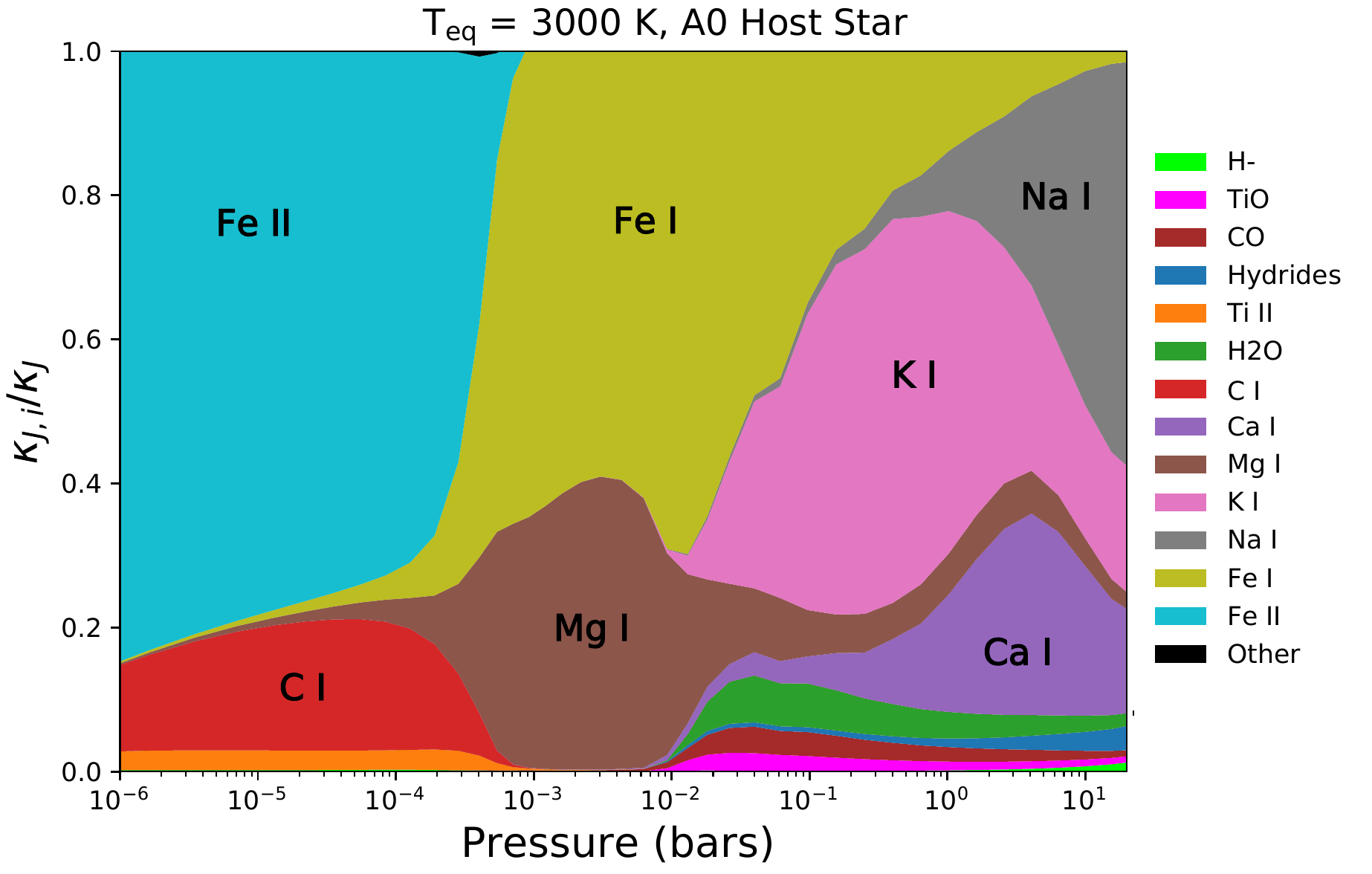}{0.51\textwidth}{(b)}}
	\caption{Relative contribution of different opacity sources to the absorption mean opacity, $\kappa_{J}$, at different pressures in the atmosphere for the T$_{eq}$ = 2250 K temperature structure without TiO/VO (left) and the T$_{eq}$ = 3000 K structure with TiO/VO (right), both irradiated by an A0-type star. TiO opacity is included in the left figure to show the contribution TiO would have if it were included.}
\end{figure*}

The nature of the temperature structure with respect to host star effective 
temperature is a consequence of the greater atmospheric opacity at short wavelengths ($<$0.5 microns) compared to the opacity at longer wavelengths (0.5-1 microns) (see Figure 8 in \cite{lothringer:2018b}). Bound-free continuous
absorption combined with absorption from atomic metals and molecules like SiO
result in large short-wavelength opacity in comparison to the absorption of
molecules like TiO and VO at longer wavelengths. 
	
	To better understand the effect of the opacity on the temperature structure, we can connect the atmosphere's mean opacities to the temperature structure \citep{mihalas:1978,hubeny:2003,hubeny:2014}. Figure~\hyperlink{fig:opac}{4} shows the ratio of the absorption mean opacity and the local Planck mean opacity. The absorption mean opacity at a given pressure is given by
	
	\begin{equation}\label{eq1}
	\kappa_{J}(P) = \frac{\int_{0}^{\infty} \kappa_{\lambda}(T,P) J_{\lambda}(P) d\lambda}{\int_{0}^{\infty} J_{\lambda}(P) d\lambda},
	\end{equation}

\noindent where $\kappa_{\lambda}$ is the monochromatic true absorption coefficient added{and }$J_{\lambda}$ is the mean intensity at a given wavelength, which is made up by a contribution from the incoming stellar irradiation and the planet's own intensity from the planet. The local Planck mean opacity is given by

\begin{equation}
\kappa_{B}(P) = \frac{\int_{0}^{\infty} \kappa_{\lambda}(T,P) B_{\lambda}(T) d\lambda}{\int_{0}^{\infty} B_{\lambda}(T) d\lambda},
\end{equation}

\noindent where in this case $B_{\lambda}(T)$ is the Planck function at the local temperature in the planet's atmosphere. The criterion for radiative equilibrium in an atmosphere can be expressed as 
\begin{equation}
	\kappa_{J}J = \kappa_{B}B.
\end{equation}

The absorption mean opacity can be thought of as the global absorption efficiency with $J$ representing a pool of photons to be absorbed \citep{hubeny:2014}. On the other hand, the Planck mean opacity can be interpreted as the efficiency with which the planet can emit radiation and cool with $B$ representing a thermal pool of photons. The ratio of the two mean opacities can describe the interplay between heating in the atmosphere and the cooling of the atmosphere through emission. Assuming the host star is a blackbody, one can approximate the temperature in the irradiation-dominated layers of an atmosphere as 
\begin{equation}\label{eq4}
	T \approx \left(\frac{\kappa_{J}}{\kappa_{B}}\right)^{1/4}f^{1/4}\sqrt{\frac{R_*}{D}} T_{*} = \left(\frac{\kappa_{J}}{\kappa_{B}}\right)^{1/4} T_{eq},
\end{equation}

\noindent where $D$ is the distance to the host star, $R_*$ and $T_*$ are the radius and temperature of the host star, respectively, and $T_{eq}$ is the equilibrium temperature \citep{hubeny:2003}. Additionally, $f$ represents a energy redistribution factor. For dayside redistribution, which we have assumed throughout this work, $f = 1/2$. For no redistribution, $f = 1$ and for full planetwide redistribution, $f = 1/4$. Note also that the term $\left(\frac{\kappa_{J}}{\kappa_{B}}\right)^{1/4}$ is equal to $\gamma$ throughout \cite{hubeny:2003}.

Equation~\ref{eq4} shows directly the relationship between the temperatures plotted in Figure~\hyperlink{fig:tps}{3} and the ratio of mean opacities plotted in Figure~\hyperlink{fig:opac}{4}. Since we have kept $T_{eq}$ constant between the cases we compare, the only difference between the models is the value of ${\kappa_{J}}/{\kappa_{B}}$ and this determines the temperature structure's behavior. The only cases where ${\kappa_{J}}/{\kappa_{B}} < 1$ are when TiO and VO are not present and $T_{eq} = 2250$~K. In these scenarios, the planet cools efficiently due to the presence of IR-active molecules like H$_2$O, while the optical opacity is not strong enough to drive an inversion. The latter effect causes ${\kappa_{J}}$ to decrease, while the former effect causes ${\kappa_{B}}$ to increase, driving the temperature lower in the upper atmosphere via equation~\ref{eq4}. In all other cases we consider, there is significant optical opacity from sources like gaseous Fe, causing ${\kappa_{J}}$ to increase. This is exacerbated by molecular dissociation, which causes ${\kappa_{B}}$ to decrease, with the net effect being a large value of ${\kappa_{J}}/{\kappa_{B}}$ and high temperatures in the irradiated layers of the atmosphere via equation~\ref{eq4}. 


We further quantify the short-wavelength opacities in ultra-hot Jupiters in Figure~\hyperlink{fig:kj}{5}, which shows the relative contribution of different opacity sources to the absorption mean opacity, $\kappa_{J}$ for two of our models. In general, we find that atomic resonance lines and doublets (e.g., Mg I 2852 \AA, Ca I 4227 \AA, Na I 5890 \& 5896 \AA, K I 7665 \& 7699 \AA) provide the largest contribution to $\kappa_{J}$ at high-pressures because of their strong line cores and pressure-broadened wings. With the exception of Mg I, these opacity sources will be hard to detect because they are present only in the deeper, isothermal layers of the atmosphere, yet they play an important role in the energy and radiative balance of the atmosphere. At lower pressures, the forest of atomic lines from Fe I, Fe II, C I, and, to a lesser-degree, Ti II provide the largest contribution to $\kappa_{J}$. Importantly, Fe I, Fe II, Mg I, and Ti II have all been detected in ultra-hot Jupiter KELT-9b \citep{hoeijmakers:2018a,cauley:2019}. Molecules like TiO, H$_2$O, and CO can also makeup a substantial portion of the total $\kappa_{J}$ until those molecules are thermally dissociated. VO does not make up a significant portion of $\kappa_{J}$ in these models.

The left-hand side of Figure~\hyperlink{fig:kj}{5} shows the makeup of the absorption mean opacity of the $T_{eq} = 2250$~K model with no TiO or VO irradiated by an A0-type host star, corresponding to the green dashed line in the left-hand panels of Figures~\hyperlink{fig:tps}{3} and \hyperlink{fig:opac}{4}. The large discontinuity at about 1 mbar is at the location of the inversion above the photosphere. Though the temperature structure for this model assumes no TiO and VO, we plot the contribution of the opacity by TiO to show TiO's ability to increase the absorption mean opacity. Even though TiO is not the dominant component of $\kappa_{J}$, TiO's addition to the model increases $\kappa_{J}/\kappa_{B}$ from 0.90 to 1.06 at 40 mbar, enough to increase the temperature by about 100 K and fundamentally change the behavior of the temperature structure. This heating would in turn affect the opacities, which would further affect the temperature structure until the model matched those with TiO and VO present. 


In the case of shallow irradiation absorption as we have here (i.e., $\kappa J_{star}\gg\kappa J_{planet}$), if one splits the absorption mean opacity into a visible and an infrared component, $\kappa_{J,vis}$ and $\kappa_{J,IR}$, respectively, it can be shown that as $\tau \xrightarrow {} \infty$,

\begin{equation}
	T_{deep} \propto \left(\frac{\kappa_{J,IR}}{\kappa_{J,vis}}+\frac{2}{3}\right)^{1/4}
\end{equation}
\noindent \citep{hansen:2008,guillot:2010}, illustrating the anti-greenhouse effect mentioned above. It can be seen that the ultra-hot Jupiters with the larger optical opacity will have a lower deep temperature.

\subsection{Stellar Spectra versus Blackbodies}\label{section:bb_tp}

\begin{figure}
    \centering
    \includegraphics[width=3.5in]{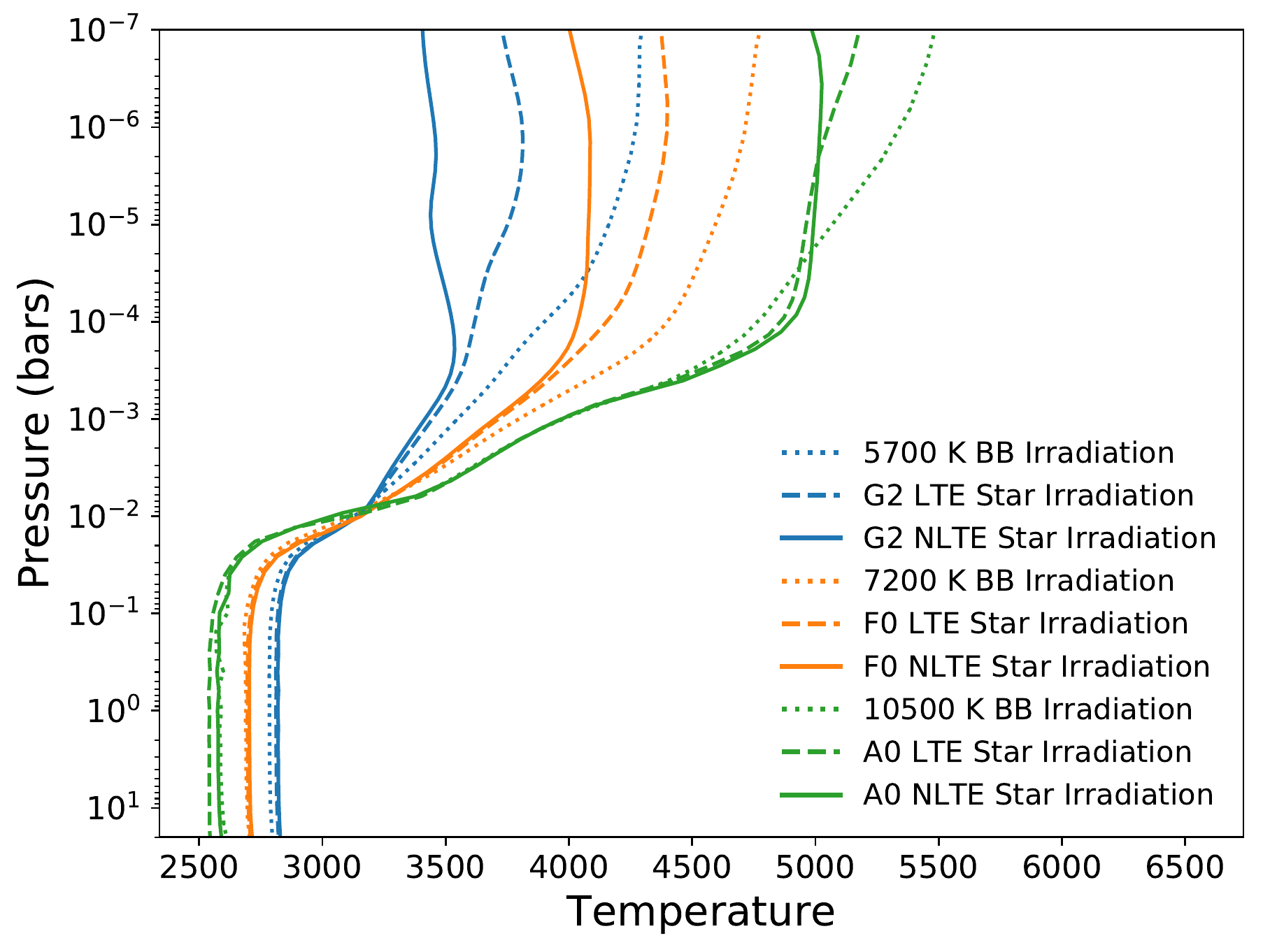}
    \caption{The temperature structure of the T$_{eq}$ = 3000 K ultra-hot Jupiters with TiO from Figure~\hyperlink{fig:tps}{3} irradiated by blackbodies (dotted), LTE stellar spectra (dashed), and the NLTE stellar spectra (solid) shown in Figure~\ref{fig:stars} and \ref{fig:nltestar}.}
    \label{fig:bb_nlte_tp}
\end{figure}

To investigate the effect of the non-blackbody nature of the stellar spectrum on the planet's temperature structure, we modeled planets irradiated by blackbodies to compare them to planets irradiated by stellar spectra. Figure~\ref{fig:bb_nlte_tp} shows that the upper atmosphere of planets irradiated by a blackbody can be much hotter than the same planet irradiated by a stellar spectrum. One cause of this increased heating is that the Planck function has more short-wavelength flux than actual stellar spectra (see Figure~\ref{fig:stars}). As we discuss above, the upper atmosphere is more opaque to this short-wavelength radiation, so even though the same amount of integrated stellar flux is absorbed by the planet, more of that flux is being absorbed at lower pressures. Similarly, since the blackbody does not have any spectral lines, there is more flux coming from the blackbody at wavelengths that species in the planet's atmosphere absorb, driving more heating.

While the planets irradiated by T$_{eff}$~=~5700 and 7200~K spectra behave similarly, the planet irradiated by T$_{eff}$~=~10500~K spectra exhibits somewhat different behavior. Both the planet irradiated by the 10500~K blackbody and the planet irradiated by the 10500~K stellar spectrum are nearly identical in temperature at pressures above 1 mbar. Between 0.1 and 1 mbar, however, the planet irradiated by the stellar spectrum is actually hotter than the planet irradiated by the blackbody spectrum. We attribute this to the greater FUV flux (1220-2000 \AA) from the star compared to the blackbody, heating the planet's atmosphere in this region. The flux from the A0 star is very high at these wavelengths because it is longward of the Lyman bound-free absorption, but shortward of the greatest Balmer bound-free absorption.

\subsubsection{The Importance of Stellar Non-Local Thermodynamic Equilibrium}\label{section:nlte_tp}

As mentioned in Section~\ref{methods:stars}, the treatment of atomic lines of a host star in LTE versus NLTE may affect the irradiated planet. Specifically, NLTE effects on the line depth of atomic metals will be important for ultra-hot Jupiters because absorption by atomic metals in particular can significantly heat an ultra-hot Jupiter's atmosphere \citep[][see Section \ref{opactemp}]{lothringer:2018b}. In Figure \ref{fig:bb_nlte_tp}, we show the temperature structure of a T$_{eq}$ = 3000 K ultra-hot Jupiter irradiated by LTE and NLTE host stars. 

We find that the planet atmosphere at pressures below 1 mbar is significantly affected by the NLTE treatment of the host star, while the remainder of the atmosphere remains relatively unaffected. Due to the deeper metal absorption lines in the NLTE star, less flux is absorbed by the atomic metals to heat the middle atmosphere, leading to temperatures  a few hundred Kelvin less at 1 microbar than the case with a star in LTE. Additionally, the effect is most important for the F0 and G2 stars. We attribute this to the fact that the metal lines in these cooler stars are deeper than in the hot A0 star combined with the fact that the cooler stars will have more neutral metals, which will match the neutral metals in the planet. At the pressures in the planet's atmosphere where the NLTE treatment of the star is relevant ($<$1 mbar), it is likely that NLTE effects in the \textit{planet's} atmosphere are also important \citep{barman:2002}. We investigate NLTE planet atmospheres in future work.

A planet irradiated by a star in NLTE but with the temperature structure of an LTE star also showed a cooler upper atmosphere, confirming that the NLTE treatment of the stellar level populations is affecting the planet's atmosphere and not changes in the stellar atmosphere's temperature structure from the NLTE treatment. Similarly, a planet irradiated with the spectrum from a stellar atmosphere that did not have Fe treated in NLTE, but had other atoms in NLTE showed a much warmer upper atmosphere. This confirmed that it is specifically the NLTE nature of atomic metals in the stellar atmosphere that is affecting the upper atmosphere of the planet.

\subsection{Atmospheric Composition} \label{results:comp}

\begin{figure*}[ht]
	\center    
	\includegraphics[width=6in]{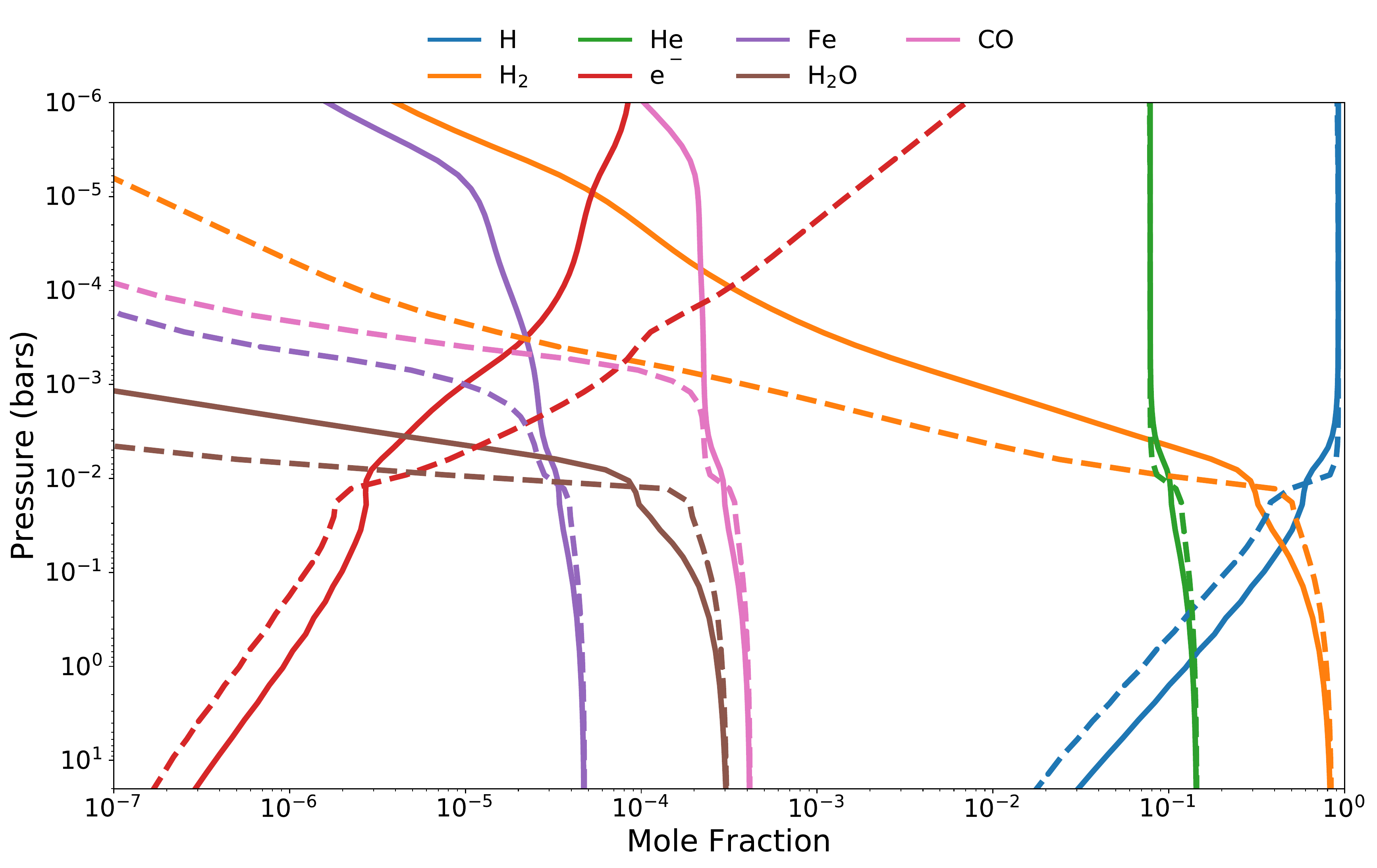}
	\caption{The mole fraction of different species in two ultra-hot Jupiters. The solid line is the T$_{eq}$ = 3000 K planet irradiated by a G2 star (T$_{eff}$ = 5700 K). The dashed line is the T$_{eq}$ = 3000 K planet irradiated by a A0 star (T$_{eff}$ = 10500 K star). \label{fig:pps}}
\end{figure*}

The planet's temperature structure has important consequences for the atmosphere's composition. Figure \ref{fig:pps} shows the mole fractions of different species in the atmosphere of a hot Jupiter with T$_{eq}$ = 3000 K irradiated by the G2 and A0 stars. As has been found previously, thermal dissociation of molecules can drastically change the composition of a planet's atmosphere from mostly molecule-dominated at higher pressures (0.1-10 bars) to mostly dominated by atoms at lower pressures (1$0^{-4}$-0.1 bars) \citep{lothringer:2018b,parmentier:2018,arcangeli:2018,kitzmann:2018}. At even lower pressures, high temperatures cause atoms to thermally ionize, which will only be exacerbated by photoionization.

The composition of an ultra-hot Jupiter's atmosphere will be affected by its host star in a number of ways. Because ultra-hot Jupiters around earlier-type, hotter host stars have steeper inversions when compared to ultra-hot Jupiters around later-type, cooler stars, thermal dissociation and ionization cause the abundance of molecules and neutral species to decrease more rapidly with decreasing pressure. For example, the mole fraction of H$_2$O will be about two orders of magnitude greater in the ultra-hot Jupiter around the G2 star than in the ultra-hot Jupiter around the A0 star at pressures of a few mbar. Similarly, specific molecular species will begin dissociating at different pressures depending on the stellar-type of the star they are orbiting. CO begins thermally dissociating around 1 mbar in the planet with the KELT-9-like A0 host star, while CO remains abundant up to 1 microbar in the planet orbiting the Sun-like G2 star. 

At pressures lower than 10 mbar, the $e^{-}$ density becomes much greater for ultra-hot Jupiters around hotter host stars than for similar planets around cooler stars due to the increased amount of thermal ionization. Correspondingly, metals like Fe begin ionizing  around 1 mbar in the planet around the A0 host star, while Fe remains neutral down to 1 microbar in the planet orbiting the G2 host star. Interestingly, the relatively cool deep temperatures of planets around hotter host stars result in slightly more molecules and neutral species existing at pressures higher than 10 mbar when compared to ultra-hot Jupiters around cooler host stars.

\section{Discussion}\label{section:discuss}

\subsection{Comparison with \cite{molliere:2015}}

\cite{molliere:2015} also looked at the effect of different host star irradiation on a hot Jupiter's atmosphere. They found a trend opposite of what we find for the hottest planets, namely that the planets around hotter host stars exhibit cooler upper atmospheres and warmer deeper atmospheres compared to planets around cooler host stars. This difference can be explained by the fact that fewer short-wavelength opacities were included in their model. Therefore, much of the incoming short-wavelength irradiation reaches the deep atmosphere and is not absorbed in the middle atmosphere where it can drive the temperature inversions in ultra-hot Jupiters. When we only consider the optical opacities relevant for planets at cooler temperatures, our models more closely match the trend described in \cite{molliere:2015}.

The statement in \cite{molliere:2015} that cooler host stars result in a more isothermal atmosphere since the absorbed radiation field becomes more similar to the emitted radiation field remains true in our models as well, since it is a consequence of Kirchoff's law of thermal radiation. The difference in our models, however, is that the ultra-hot Jupiters around cooler host stars are more isothermal because they have less steep thermal inversions in contrast to planets around hotter host stars. 

Another way to describe this behavior is by considering the mean opacities described in Section~\ref{results:tps}. As the effective temperature of the host star approaches the atmospheric temperature, ${\kappa_{J}} \xrightarrow{}{\kappa_{B}}$ and ${\kappa_{J}}/{\kappa_{B}} \xrightarrow{} 1$. If this hold throughout the highly irradiated layers where equation~\ref{eq4} holds, then the atmosphere is isothermal with $T \xrightarrow{} T_{eq}$ in this region. With ${\kappa_{J}}/{\kappa_{B}} = 1$ there is neither greenhouse, nor anti-greenhouse effects, resulting in an isothermal atmosphere.

\subsection{Observational Implications}\label{discuss:obs}

\begin{figure*}[ht!]
	\gridline{\fig{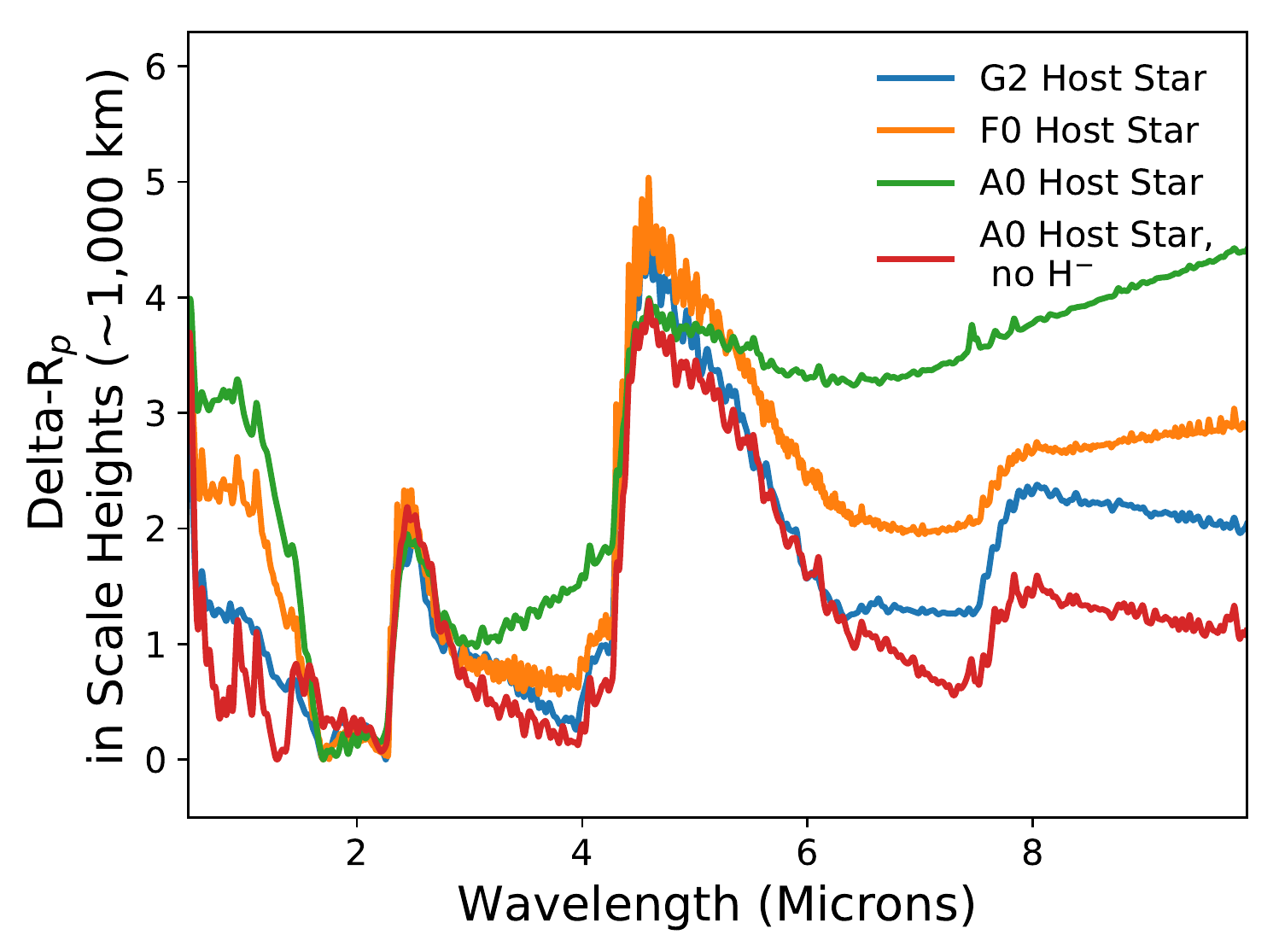}{0.5\textwidth}{(a)}\hypertarget{fig:obsa}{}}
	\gridline{\fig{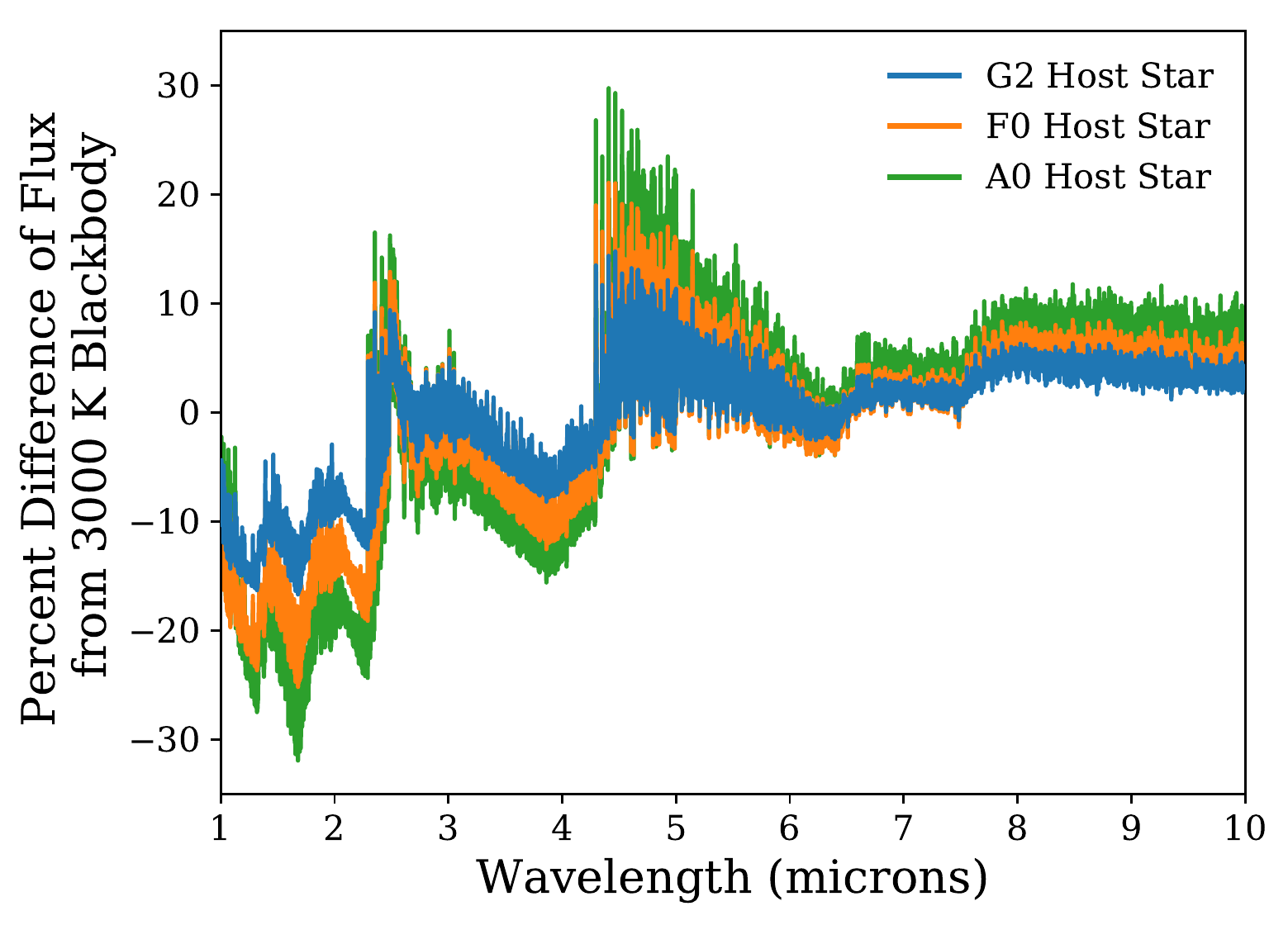}{0.5\textwidth}{(b)}\hypertarget{fig:obsb}{} 
		\fig{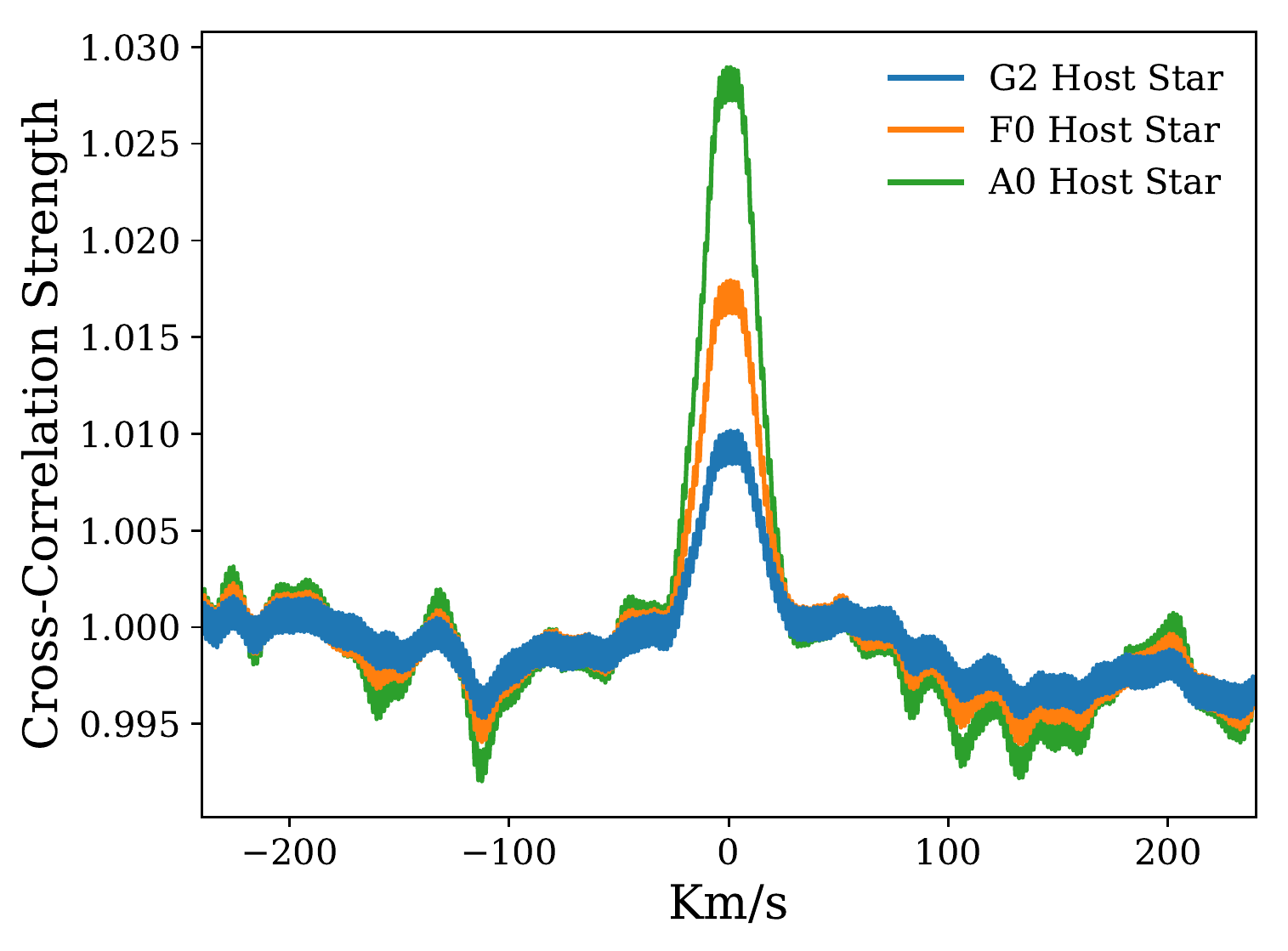}{0.5\textwidth}{(c)} \hypertarget{fig:obsc}{}}
			
	\caption{Observational implications of host star irradiation. Panel (a) shows the transit spectrum in scale heights above the minimum near-IR transit radius of of the T$_{eq}$ = 3000K planet around different host star types. The ultra-hot Jupiters around hotter host stars have a muted transit spectrum due primarily to increased H$^{-}$ opacity in comparison to planets around cooler stars. Panel (b) shows the emission spectrum of the same planet in percent difference from a 3000 K blackbody. Planets around hotter host stars have larger spectral features due to the increased contrast in brightness temperature between pressures probed at these wavelengths. Panel (c) shows the result of cross-correlating the high resolution emission spectrum of the same planet between 2.3 and 2.6 microns with a CO template binary mask, resulting in the mean CO line profile. Ultra-hot Jupiters around hotter stars have greater line-to-continuum contrasts compared to similar planets around cooler stars for the same reason the lower resolution emission spectrum has larger spectral features.}\label{fig:obs}
\end{figure*}

Ultra-hot Jupiters are ideal targets to observe due to their high temperature, inflated radii, and short periods. To properly interpret such observations, it is important to understand the observational implications that the planet's host star dependencies will have. Figure \ref{fig:obs} shows how the planet's (a) transit spectrum, (b) emission spectrum, and (c) CO line contrast vary with host star irradiation. These three quantities are observable during transit, secondary eclipse, and in high-dispersion spectroscopy, respectively. We focus on CO since it will be the best molecular probe of ultra-hot Jupiter atmospheres \citep{lothringer:2018b,kitzmann:2018}. 

Figure~\hyperlink{fig:obsa}{7a} demonstrates that ultra-hot Jupiters around hotter stars show much more muted spectral features in transit than similar planets around cooler stars. This effect is primarily a result of the increased H$^{-}$ continuous opacity in hotter planets, which raises the infrared photosphere and results in weaker CO spectral features, but is exacerbated by the increased thermal dissociation of CO at pressures below 1 mbar. This behavior is demonstrated by the red model in Figure~\hyperlink{fig:obsa}{7a}, which shows large CO features in the transit spectrum in the absence of H$^{-}$ opacity. For planets of the same equilibrium temperature, it is therefore more advantageous to observe ultra-hot Jupiters around cooler host stars rather than ultra-hot Jupiters orbiting hot host stars when targeting molecular bands in transit.

H$^{-}$ opacity is not gray, but rather increases toward both short and long wavelengths from its minimum at 1.6 microns, causing the transit radius to exhibit a negative slope shortward of 1.6 microns, and then a positive slope longward of 1.6 microns. The slope is steeper for ultra-hot Jupiters around hot host stars because of the increased H$^{-}$ abundance. Such a slope is detectable with future observations with the infrared coverage of JWST. Note that the transit spectrum is calculated with the temperature structures from Figure~\hyperlink{fig:tps}{3} for consistency. The terminator on such planets is likely cooler than the dayside temperatures, but the trends with host star irradiation spectrum are still applicable until the atmosphere reaches temperatures below about 2500 K, where H$^{-}$ opacity becomes weak.

The effect for the planet's emission spectrum, seen in Figure~\hyperlink{fig:obsb}{7b}, is the opposite of the transit spectrum: ultra-hot Jupiters around hotter stars have larger CO features than ultra-hot Jupiters around cooler stars. The large CO features are a consequence of the wide range of brightness temperatures probed by emission spectra in planets with steeper inversions. Similar to the transit spectrum, however, are the slopes caused by H$^{-}$ opacity. The slope of the emission spectrum increases with host star effective temperature because of the increasing contrast in brightness temperature probed by H$^{-}$ opacity.

\begin{figure*}[ht!]
	\gridline{\fig{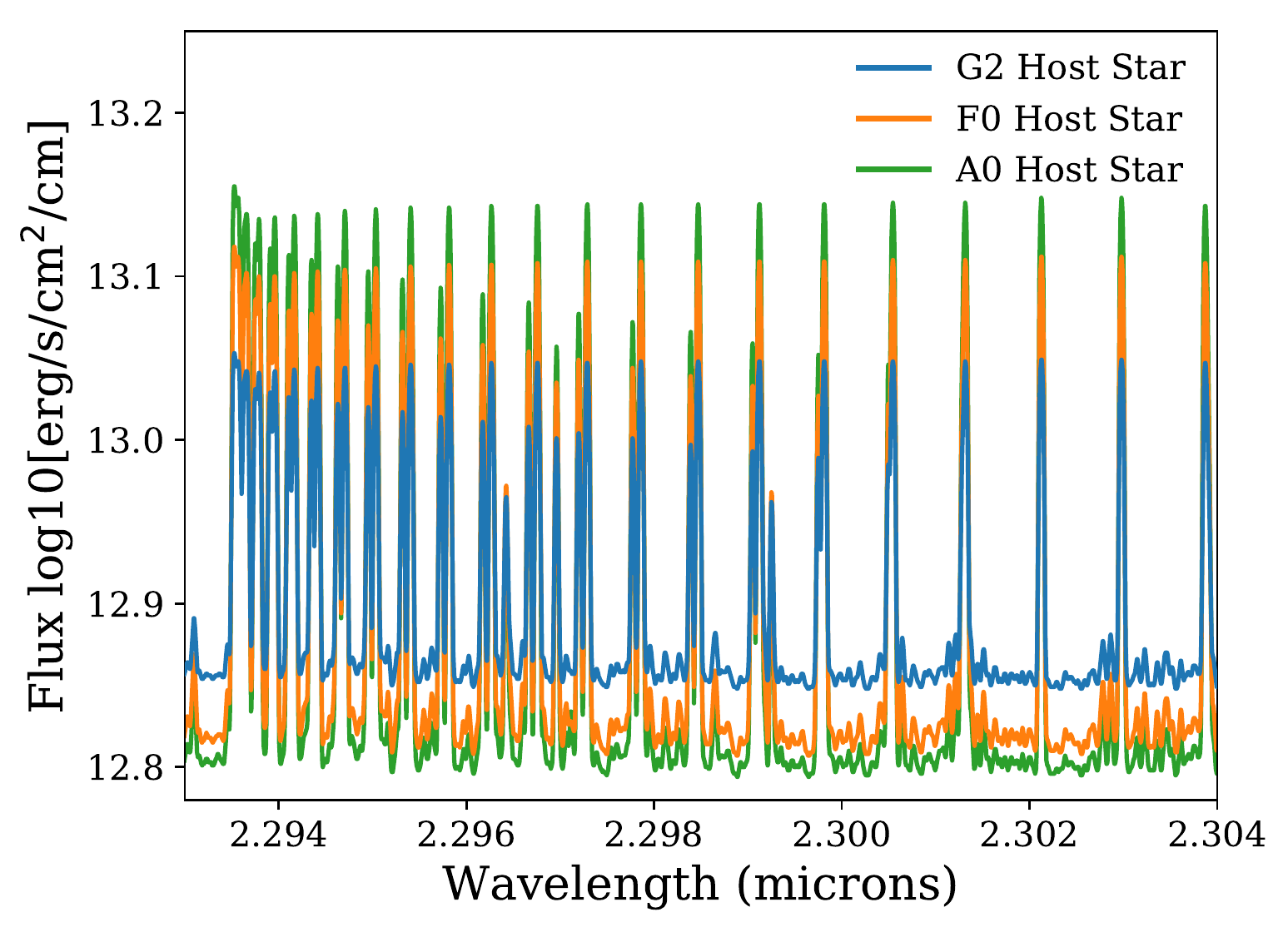}{0.5\textwidth}{(a)} 
		\fig{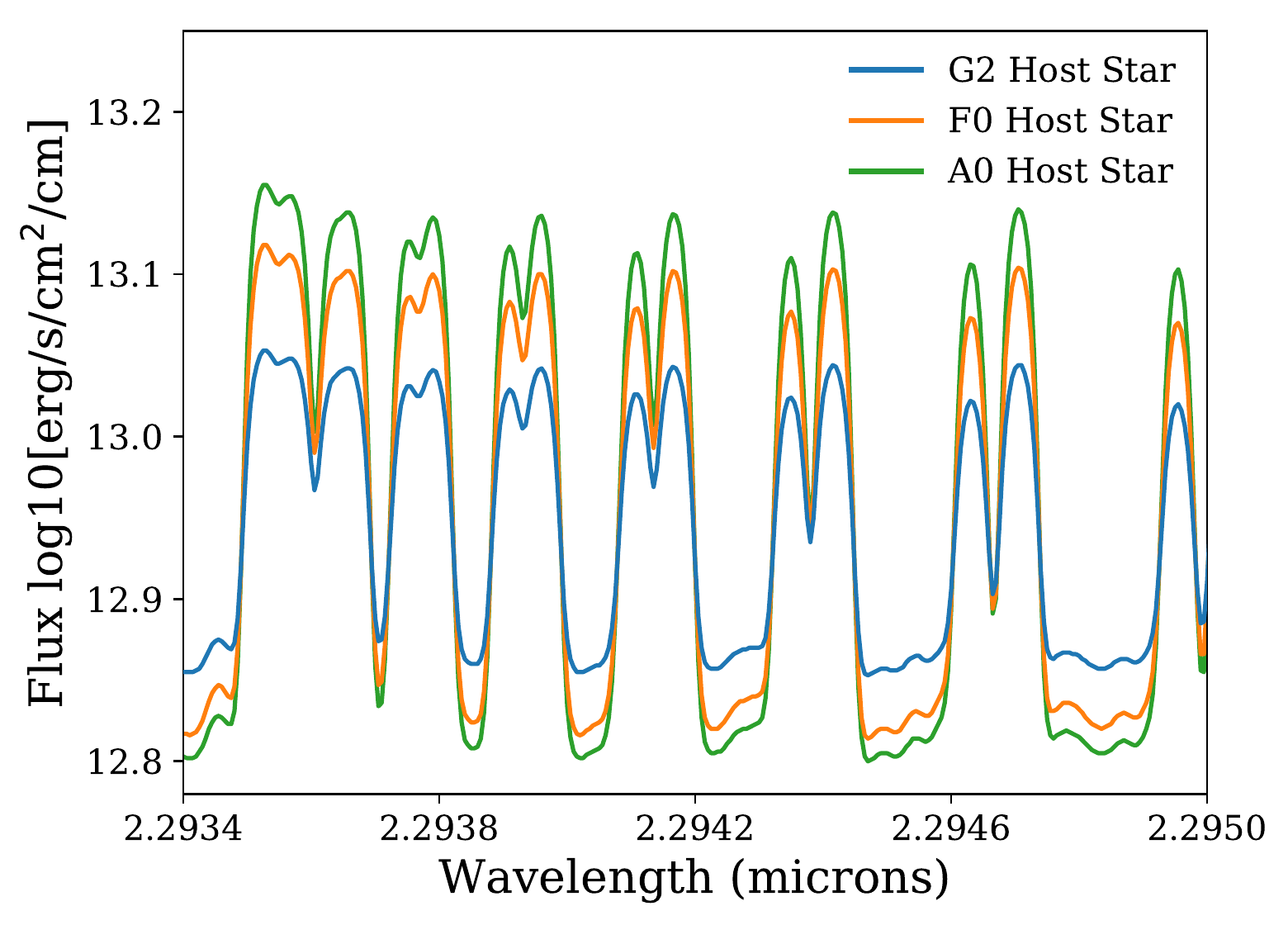}{0.5\textwidth}{(b)} }
			
	\caption{High-resolution (R$\sim$500,000) spectra of the T$_{eq}$ = 3000 K hot Jupiters, focusing on the first overtone band of CO at 2.3 microns. The figure at right zooms in on the figure at left. 
	}\label{fig:obs_hires}
\end{figure*}

The increased contrast in brightness temperature for planets around hotter stars can also be explored with high-dispersion spectroscopy (HDS). After a planet's spectrum is observed with HDS and telluric lines are removed, the continuum-normalized spectrum is often cross-correlated with either a template from a model atmosphere or a binary mask of line positions. When a mask is used, the resulting function is the mean line profile \citep[e.g.,][]{pino:2018}. Figure~\hyperlink{fig:obsc}{7c} shows this mean line profile for the CO lines between 2.3 and 2.6 microns. As with the planet's emission spectrum, the line contrasts are greater in planets with steeper inversions (i.e., those around hotter host stars). This result is also demonstrated in Figure~\ref{fig:obs_hires}, which shows individual CO emission lines at R$\sim$500,000. 

A disadvantage to targeting bright host stars, however, is the host star's effect on the signal-to-noise of the desired planetary signal. The secondary eclipse signal-to-noise scales as 

\begin{equation}
SNR = \frac{F_{p}}{\sqrt{F_{s}}} 
\end{equation}	
	
\noindent where $F_{p}$ is the flux from the planet and $F_{s}$ is the flux from the star \citep{tessenyi:2013}. If we assume each host star has the same apparent magnitude, then hotter and brighter stars will be farther away and the flux from planets around them will be correspondingly lower. Even though the CO signal increases by about 10\% between the ultra-hot Jupiter around an A0-type star compared to an ultra-hot Jupiter around a G2-type star, the SNR of the secondary eclipses will decrease due to the decreased flux from the planet because the A0 system will be $\sim 3.4x$ farther away than the G2 system.
	
%
%

From \cite{molliere:2018}, the HDS signal-to-noise is given by

\begin{equation}
	SNR = \frac{1}{\sqrt{F_s}} \left(\sum_{i=1}^{N} I_i^2\right)^{1/2}
\end{equation} 

\noindent where $N$ is the number of lines measured, $I_i$ is the strength of line $i$, and $F_s$ is the stellar flux, which we've assumed is equal in magnitude between the cases and the dominant source of noise. Again, even though the strength of the lines probed in an ultra-hot Jupiter around an A0-type star is $2.5x$ larger than the same line in an ultra-hot Jupiter around a G2-type star, the fact that the planet is $3.4x$ father away and therefore $11.5x$ less bright means that the SNR of CO observations will be greater in an ultra-hot Jupiter around a G2-type star compared to a A0-type star.


	
	
	

\subsection{Consequences for Magnetohydrodynamics}

Global circulation models have shown that drag from magnetic Lorentz forces has the ability to damp a hot Jupiter atmosphere's circulation by decreasing zonal wind speeds, thus modifying the day-night temperature contrast for tidally locked planets \citep{perna:2010a,rauscher:2012,batygin:2013,rogersandtad:2014,rogersandshowman:2014,tad:2016}. The magnitude of these drag forces depends both on the ion fraction, as well as the strength of the magnetic field. The former of these, the ion fraction, will itself depend on the atmosphere's temperature since the main source of ions in the lower and middle atmosphere will be from thermal ionization. 

In Section \ref{results:comp}, we showed that the ion fraction in an ultra-hot Jupiter's atmosphere will depend on the irradiation spectrum from the planet's host star. The Lorentz forces that potentially control the drag in an atmosphere's circulation will therefore also depend on the stellar type of the host star. In the limit of large drag forces, $V_{\phi} \propto \sqrt{T}/\chi_{e^{-}}$, where $V_{\phi}$ is the longitudinal wind speeds and $\chi_{e^{-}}$ is the ion fraction \citep{menou:2012b}. Ultra-hot Jupiters around hotter host stars (i.e., with steeper and hotter temperature inversions) will have a higher ion fraction in the middle and upper atmosphere and a correspondingly larger drag force. At 1 mbar, $\chi_{e^{-}}$ is about an order of magnitude larger for an ultra-hot Jupiter around a A0 star than a planet around a G2 star, while the temperature is about 1000 K larger. This corresponds to a wind speed that is a factor of ${\sim}{9}$ slower in the planet around a A0 star. However, deeper in the atmosphere, where the radiative timescale is longer, ultra-hot Jupiters around hotter host stars can be a few hundred Kelvin cooler, resulting in less thermal ionization and potentially a smaller drag force. Using the scaling law above, the wind speeds at 1 bar in the planet around a G2-star should be about twice as fast as the planet around a A0 star. 

Additionally, previous studies have suggested that atmospheres with temperature inversions likely have larger day-night temperature contrasts due to the fact that the stellar irradiation is absorbed at lower pressures where the radiative timescale is shorter \citep{fortney:2008,dobbs-dixon:2008,perna:2012}. Ultra-hot Jupiters around hotter host stars may have even greater day-night temperature contrasts due to their steeper and hotter temperature inversions in comparison to ultra-hot Jupiters around cooler stars. 

Lastly, recent modeling has shown that the hottest Jovian planets may exhibit atmospheric dynamos \citep{rogers:2017b}, which may be enhanced in ultra-hot Jupiters around hotter host stars. Such an enhancement would have consequences for the planet's magnetic field strength and geometry, as well as potential planet-star interactions. Further GCM modeling can shed light on these phenomena and future observations of ultra-hot Jupiter phase curves can confirm trends with host star type.

\section{Conclusions}

We have shown that the basic structure, composition, observed spectra, and likely the circulation of an ultra-hot Jupiter will depend on the properties of the spectrum irradiating the planet. At the same equilibrium temperature, ultra-hot Jupiters around early-type, hot host stars will have steeper temperature inversions and greater maximum temperatures than planets orbiting cooler stars. The amount of short-wavelength absorption determines the steepness and magnitude of the temperature inversion by driving strong heating at milibar pressures and ultra-hot Jupiters around hot host stars experience more short-wavelength irradiation ($<$0.5 microns) compared to similar planets around cooler stars. This steeper and hotter temperature inversion in planets around hot host stars, in contrast to planets around cooler host stars, will have several important consequences:
\begin{itemize}
    \item Thermal dissociation of molecules at low pressures will more significantly decrease molecular abundances in ultra-hot Jupiters around hotter stars.
    \item The increased amount of H$^{-}$ opacity in ultra-hot Jupiters around hotter stars will mute spectral features in the transit spectrum.
    \item The large brightness temperature contrasts resulting from steep temperature inversions enhance spectral features and line contrasts in the emission spectrum of planets around hotter stars.
    \item The larger ion fraction and increased radiation absorption at low pressures in ultra-hot Jupiters around hotter stars may modify the atmosphere's circulation and increase day-night temperature contrasts.
\end{itemize}

	
We have also quantified the opacity sources responsible for the absorption of irradiation and the heating of the atmosphere. Atomic resonance lines from Mg I, Ca I, Na I, and K I, combined with molecular opacity from TiO, H$_2$O, and CO provide the major opacities in the atmosphere at high pressures, while the forest of lines from species like Fe I and Fe II dominate at lower pressures.

TESS and the continuing efforts of ground-based surveys like WASP and KELT will find more ultra-hot Jupiters around a wide range of host star types, including many early-type stars \citep{barclay:2018}. In order to fully understand these planets, we need to understand the effect that different host stars have on their planet's atmosphere. To properly interpret observations of highly irradiated planets, we must be aware of their host stars and have a detailed understanding of the irradiation spectrum, including its NLTE properties. The trends we predict here will be readily detectable with further observations of ultra-hot Jupiters from both the ground with HDS and from space with HST, \textit{Spitzer}, and JWST. These predicted trends will also help observers prioritize ultra-hot Jupiter targets. 
 
\acknowledgments
We thank the anonymous referee for helpful comments and attention to detail that improved the manuscript. This research was partially supported under programs HST-GO-12511 and HST-GO-14797, with financial support provided by NASA through a grant from the Space Telescope Science Institute, which is operated by the Association of Universities for Research in Astronomy, Inc., under NASA contract NAS 5-26555. This research has made use of the NASA Astrophysics Data System and the NASA Exoplanet Archive, which is operated by the California Institute of Technology, under contract with the National Aeronautics and Space Administration under the Exoplanet Exploration Program. Allocation of computer time from the UA Research Computing High Performance Computing (HPC) at the University of Arizona is also gratefully acknowledged. 

\software{PHOENIX \citep{hauschildt:1997,barman:2001}}

\bibliographystyle{aasjournal}

\end{document}